\documentclass[12pt]{article}
\pdfoutput=1    
  
\usepackage{hyperref}    
\usepackage{amsmath}      
\usepackage{amssymb} 
\usepackage{graphicx}
\usepackage{url}
\usepackage{enumerate}
 
\usepackage{float,wrapfig,color} 

\allowdisplaybreaks
\def\eq#1{(\ref{#1})}
\def\s[#1\s]{\begin{align}\begin{split}#1\end{split}\end{align}}
\def\[#1\]{\begin{align}#1\end{align}}
\def\bpsi{{\bar \psi}}
\def\sig#1{{\sigma_#1}}
\def\pbpsi{\bpsi_\parallel}
\def\ppsi{\psi_\parallel}
\def\tbpsi{\bpsi_\perp}
\def\tpsi{\psi_\perp}

\def\teta{\tilde \eta}

\begin{document}

\begin{titlepage} 

\title{
\hfill\parbox{4cm}{ \normalsize YITP-22-85}\\  
\vspace{1cm} 
Signed distributions of real tensor eigenvectors of \\ Gaussian tensor model via a four-fermi theory}

\author{Naoki Sasakura\footnote{sasakura@yukawa.kyoto-u.ac.jp}
\\
{\small{\it Yukawa Institute for Theoretical Physics, Kyoto University, }}\\
{\small {\it and } } \\
{\small{\it CGPQI, Yukawa Institute for Theoretical Physics, Kyoto University,}} \\
{\small{\it Kitashirakawa, Sakyo-ku, Kyoto 606-8502, Japan}}
}


\maketitle 

\begin{abstract}  
Eigenvalue distributions are important dynamical quantities in matrix models, and 
it is a challenging problem to derive them in tensor models.  
In this paper, we consider real symmetric order-three tensors with Gaussian distributions as the simplest case, 
and derive an explicit formula for signed distributions of real tensor eigenvectors: 
Each real tensor eigenvector contributes to the distribution by $\pm 1$, 
depending on the sign of the determinant of an associated Hessian matrix. 
The formula is expressed by the confluent hypergeometric function of the second kind, which is obtained by computing 
a partition function of a four-fermi theory.
The formula can also serve as lower bounds of real eigenvector distributions (with no signs), and
their tightness/looseness are discussed by comparing with Monte Carlo simulations.
Large-$N$ limits are taken with the characteristic oscillatory behavior of the formula being preserved. 
\end{abstract}
\end{titlepage}

\section{Introduction}
\label{sec:intro}
Eigenvalue distributions are one of the main tools in computations of matrix models \cite{Wigner,Brezin:1977sv}, 
and are also useful for qualitative understanding of the dynamics through their topological 
properties \cite{Gross:1980he,Wadia:1980cp}. 
It would be an interesting problem to study similar distributions in tensor models
\cite{Ambjorn:1990ge,Sasakura:1990fs,Godfrey:1990dt,Gurau:2009tw}. 

Though it is not difficult to numerically compute eigenvalues/vectors for a given tensor of a small size by using  
commonly-used computers, analytical understanding of their properties and large-$N$ limits (thermodynamic limits) for ensembles 
of tensors are still very limited: In \cite{realnum1,realnum2} the expected numbers of real tensor eigenvalues are computed;
In \cite{Evnin:2020ddw} the largest eigenvalue of a typical tensor in a Gaussian ensemble is estimated; 
in \cite{Gurau:2020ehg} the Wigner semicircle law in matrix models is extended to tensor models.
In this paper, we give an explicit formula for real tensor eigenvector distributions with signs for the simplest case. 
The formula is derived by computing a partition function of a four-fermi theory.

As the simplest case, we restrict ourselves to a real symmetric tensor of
order-three, $C_{abc}\ (C_{abc}=C_{bac}=C_{bca} \in \mathbb{R},\ a,b,c=1,2,\ldots,N)$.  
There exist various definitions of tensor eigenvalues/vectors \cite{Qi,lim,cart}. In this paper, we employ a definition
of real eigenvectors,
\[
C_{abc} v_b v_c=v_a, \ v\neq 0,\ v\in\mathbb{R}^N.
\label{eq:cv}
\]
Note that repeated indices are assumed to be summed over throughout
this paper. 
Real eigenvalues $h$ accompanied with real eigenvectors (Z-eigenvalues in \cite{Qi}) can be deduced by
normalizing $v$ as $w=v/|v|$,
\[
C_{abc} w_b w_c=h \, w_a, \ |w|:=\sqrt{w_a w_a}=1,
\label{eq:egw}
\]
with 
\[
h=1/|v|.
\label{eq:relation}
\]

The distribution of $v_a$ for each case of $C_{abc}$ is given by
\def\const{\hbox{const.}}
\[
\rho(C,v)=\left | \det M (v)\right| \prod_{a=1}^N \delta \left(v_a - C_{abc}v_b v_c \right),
\]
where the matrix $M(v)$ has components,
\[
M(v)_{ab}=  \delta_{ab} -2 C_{abc} v_c,
\]
and $\det$ denotes the matrix determinant. 
The determinant factor is to make 
each solution identically contribute under the measure $dv=\prod_{a=1}^N dv_a$.
In fact,
\[
\rho(C,v)=\sum_{i=1}^{\#{\rm real\ sol.}(C)} \delta^N (v-v^i),
\label{eq:rhosum}
\]
because of $M(v)_{ab}=\frac{\partial}{\partial v_b} (v_a - C_{acd}v_c v_d)$,
where $v^i\ (i=1,2,\ldots, \#\hbox{real sol.($C$}))$ denote all the real solutions to \eq{eq:cv} for a given $C$.

The eigenvector equation \eq{eq:cv} can be considered to be a stationary point equation of a potential 
$V=v_a v_a /2- C_{abc}v_a v_b v_c/3$. Then the matrix $M(v)$ is a Hessian matrix at
the stationary point.

An interesting quantity is the mean distribution of $v$ under a Gaussian distribution of $C$:
\[
\rho(v)=\langle \rho(C,v) \rangle_C=A^{-1} \int dC \exp(-\alpha\, C^2)   \left | \det M(v) \right | 
\prod_{a=1}^N \delta \left(v_a - C_{abc}v_b v_c \right),
\label{eq:defrho}
\]  
where $C^2=C_{abc}C_{abc}$, $\alpha$ is a positive constant, and $A=\int dC \exp(-\alpha \, C^2)$ is a normalization factor. 
Since the integration over $C$ is $O(N)$ symmetric, $\rho(v)$
actually depends only on $|v|=\sqrt{v_a v_a}$.

It would not be straightforward to compute \eq{eq:defrho}. We rather consider a more tractable quantity,
\def\trho{{\tilde \rho}}
\[
\trho(v)=A^{-1} \int dC \exp(-\alpha\, C^2)    \det M(v)  
\prod_{a=1}^N \delta \left(v_a - C_{abc}v_b v_c \right),
\label{eq:trho}
\]
in the rest of this paper. The difference from \eq{eq:defrho} is that taking the absolute value has been ignored. 
Therefore, $\trho(v)$ is the quantity,
\[
\trho(v)=\left\langle 
\sum_{i=1}^{\#{\rm real\ sol.}(C)} \hbox{sign}(M(v^i))\, \delta^N (v-v^i)
\right\rangle_C,
\]
with an additional sign factor compared to \eq{eq:rhosum}.
We may call it a signed eigenvector distribution because of this sign factor.

There do not seem to exist any apparent quantitative relations between $\rho$ and $\trho$, but 
$\trho$ can be used as a lower bound of $\rho$,  since
\[
| \trho(v) |\leq \rho(v).
\label{eq:ineq}
\]
We can also expect a similar relation, 
\[
|\teta (h)| \leq \eta(h) ,
\]
to hold between a signed Z-eigenvalue distribution $\teta$, and a Z-eigenvalue distribution $\eta$.
Here, using the relation \eq{eq:relation} and that $\trho(v)$ depends only on $|v|$, 
the signed Z-eigenvalue distribution is defined by
\[
\teta(h)=\trho(|v|) S_{N-1} |v|^{N-1}  \left| \frac{d|v|}{dh}\right |=\trho(1/h) \, S_{N-1}\, h^{-N-1},
\]
where $S_{N-1}=2 \pi^{N/2}/\Gamma(N/2)$ is the surface area of the unit hypersphere in $N$-dimensions.

A potential mathematical interest of the signed distribution above could be found in relation with Morse theory. 
The potential $V$ above may be regarded as a random Morse function, and summations of signs over 
stationary points are related to Euler characteristics. Therefore $\tilde \rho(v)$ will reflect some topological aspect
over sections of constant $|v|$ in terms of random Morse functions.

A more direct physical application of the signed distribution can be found in the context of spin glasses.
The Hamiltonian of the spherical $p$-spin model \cite{pspin,pedestrians} with $p=3$ is defined by 
\s[
H=C_{abc} w_a w_b w_c \hbox{ with constraint } w_a w_a=1,
\label{eq:ham}
\s]
where $w$ is the dynamical variable, and $C$ is a random external field.
The Hamiltonian has in general a macroscopic number of stationary points, which is called complexity, 
and it is important to know distributions of local minimums to understand the dynamics. By implementing the 
constraint in \eq{eq:ham} by the method of Lagrange multiplier, one can find that the stationary points are given
by the solutions to \eq{eq:egw} with $\pm h$ being the energy, and the sign of $M$ 
corresponds to $(-1)^{\#{\rm negative}+1}$, where $\#\hbox{negative}$ is the number of unstable directions
around a stationary point when the energy is negative\footnote{The $+1$ of the exponent comes from the radial 
direction, which is irrelevant for the stability due to the constraint.}. Thus the signed distribution
can provide a good estimate of the energy range where local minimums dominate by looking at the negative 
value region of $\tilde \rho(v)$ near the smallest end of $|v|$.
We would also like to add that the present paper is closely related to 
the study \cite{randommat}, which counts the stationary points of the Hamiltonian of 
the spherical $p$-spin model 
in the large-$N$ (thermodynamic) limit by using random matrix theory\footnote{
In spin glass physics, the distribution of local minimums of 
the spherical $p$-spin model is rather discussed 
in terms of the TAP free energy. For example see \cite{example} for the status.}.   
    
\section{A four-fermi theory}
\eq{eq:trho} can be recast into a more tractable form by introducing some virtual variables. 
As well known, the determinant can be 
rewritten by fermionic variables by using $\int d\bpsi d\psi \, e^{\bpsi_a K_{ab} \psi_b} =\det K$ \cite{zinn}, 
and the delta function by a bosonic variable:
\[
\trho(v)=(2\pi)^{-N} A^{-1} \int dC d\bpsi d\psi d\lambda \, e^{S},
\label{eq:trhowiths}
\]
where the action $S$ is given by
\[
S=-\alpha \, C^2 +i \lambda_a \left(v_a - C_{abc}v_b v_c \right)+\bpsi_a  \left( \delta_{ab} -2 C_{abc} v_c\right) \psi_b,
\label{eq:s}
\]
and $\lambda$ and $\bpsi,\psi$ are respectively bosonic and fermionic.
Below we will integrate over the bosonic variables to finally obtain a fermionic theory,
assuming that this change of the order of the integrations does not affect the final result.

Let us first integrate over $C$. The part containing $C$ in \eq{eq:s} is given by
\[
-\alpha\, C^2-2 C_{abc} \bpsi_a \psi_b v_c-i \, C_{abc} \lambda_a v_b v_c.
\]
Performing the Gaussian integration over $C$ results in the cancellation of the prefactor $A^{-1}$ in \eq{eq:trhowiths},
and a change of the action by 
\[
\delta_C S=\frac{1}{\alpha} \left(\frac{1}{6} \sum_{\sigma} \left(\frac{i}{2} \lambda_{\sig{a}} v_{\sig{b} }v_\sig{c} +\bpsi_\sig{a} \psi_\sig{b} v_\sig{c}\right) \right)^2,
\label{eq:sc1}
\]
where $\sigma$ denotes summation over all the permutations of the indices, $a,b,c$.
Expanding the expression in \eq{eq:sc1}, we obtain
\s[
&-\frac{1}{12 \alpha} \left( \lambda^2 (v^2)^2 + 2 v^2 (v\cdot \lambda)^2 \right)
+\frac{i}{3\alpha } \left( \bpsi\cdot \lambda \, \psi\cdot v\, v^2+\bpsi\cdot v \, \psi \cdot \lambda \, v^2+\bpsi\cdot v \, \psi\cdot v \, \lambda \cdot v \right) \\
& -\frac{1}{3\alpha } \bpsi\cdot \psi \, \bpsi\cdot v \, \psi\cdot v-\frac{1}{6\alpha } (\bpsi\cdot \psi)^2 v^2,
\label{eq:resintc}
\s]
where $A\cdot B=A_a B_a$. To derive this expression, we have used $\bpsi\cdot \bpsi=\psi\cdot \psi=0$,
which follow from the fermionic property.

Let us next perform the integration over $\lambda$. Picking up the terms containing $\lambda$ in \eq{eq:s} (with no $C$) 
and \eq{eq:resintc}, we have
\[
-\frac{1}{12 \alpha} B_{ab} \lambda_a \lambda_b+ i \lambda_a D_a,
\
\]
where
\s[
B_{ab}&=(v^2)^2 \delta_{ab} + 2 v^2 v_a v_b, \\
D_a&=v_a+\frac{1}{3\alpha } \left( \bpsi_a \, \psi\cdot v\, v^2+\bpsi\cdot v \, \psi_a \, v^2+\bpsi\cdot v \, \psi\cdot v \,  v_a \right). 
\label{eq:bd}
\s]
The inverse of $B$ is straightforwardly determined to be
\[
B^{-1}_{ab}=(v^2)^{-2} \delta_{ab} -\frac{2}{3} (v^2)^{-3} v_a v_b.
\]
Therefore the change of the action by the Gaussian integration over $\lambda$ is obtained as
\s[
\delta_\lambda S&=-\frac{1}{2} \ln \det B-3 \alpha B^{-1}_{ab}D_a D_b \\
&= -\frac{1}{2} \ln \det B-3 \alpha (v^2)^{-2} D^2 +2 \alpha (v^2)^{-3} (v\cdot D)^2,
\label{eq:dellam}
\s]
and a multiplicative factor $(12  \pi \alpha)^{\frac{N}{2}}$.

The determinant of $B$ in \eq{eq:dellam} can easily be determined because the eigenvalues of $B$ are 
$3 (v^2)^2$ with no degeneracy and $(v^2)^2$ with degeneracy $N-1$, where the corresponding eigenvectors are $v$ 
and all the vectors transverse to $v$, respectively.
Thus,
\[
\ln \det B=\ln 3+2 N \ln v^2.
\label{eq:detb}
\] 
As for the second term, we obtain
\[
D^2=v^2+\frac{2}{\alpha} \bpsi\cdot v \, \psi\cdot v \, v^2-\frac{2 (v^2)^2}{9 \alpha^2} \bpsi\cdot \psi \, \bpsi\cdot v \, \psi\cdot v,
\label{eq:d2exp}
\]
after a straightforward computation shown in \ref{app:d2}.
To compute the last term in \eq{eq:dellam}, we first have
\[
v\cdot D=v^2+\frac{1}{\alpha} \bpsi\cdot v \, \psi\cdot v \, v^2.
\]
Therefore,
\[
(v\cdot D)^2=(v^2)^2+\frac{2}{\alpha} \bpsi\cdot v \, \psi\cdot v \, (v^2)^2,
\label{eq:vd2}
\]
where we have used $(\bpsi\cdot v)^2=0$ (or $(\psi\cdot v)^2=0$) because of the fermionic property.
Putting \eq{eq:detb}, \eq{eq:d2exp}, and \eq{eq:vd2} into \eq{eq:dellam}, we obtain
\[
\delta_{\lambda} S=-\frac{1}{2}\log 3 -N \ln v^2 -\frac{\alpha}{v^2} -\frac{2}{v^2} \bpsi\cdot v \, \psi\cdot v+\frac{2}{3 \alpha} \bpsi\cdot \psi\,\bpsi\cdot v\,\psi \cdot v.
\label{eq:dlams}
\]
Adding the remaining terms in \eq{eq:s} and \eq{eq:resintc} to \eq{eq:dlams}, 
and taking into account the generated multiplicative factors, we finally obtain an expression
of $\tilde \rho$ via a four-fermi theory,
\[
\trho(v)=3^\frac{N-1}{2} \pi^{-\frac{N}{2}} \alpha^{\frac{N}{2}}\int d\bpsi d\psi \, e^{S_{\bpsi\psi}},
\label{eq:trhopsi}
\]
with
\[
S_{\bpsi\psi}=-\frac{\alpha}{v^2} -N \ln v^2+\bpsi\cdot \psi -\frac{2}{v^2} \bpsi\cdot v \, \psi\cdot v-\frac{v^2}{6 \alpha} (\bpsi \cdot 
\psi)^2+\frac{1}{3 \alpha} \bpsi\cdot \psi \, \bpsi\cdot v \, \psi\cdot v.
\label{eq:fermionaction}
\]

\section{An explicit expression of $\trho(v)$}
To further compute \eq{eq:trhopsi}, it is more convenient to separate $\bpsi,\psi$ into the parallel and transverse directions against $v$:
\s[
\bpsi&=\pbpsi+\tbpsi, \\
\psi&=\ppsi+\tpsi,
\s]
where $\tbpsi\cdot v=\tpsi\cdot v=0$, $\pbpsi\cdot v=|v|\, \pbpsi$, $\ppsi \cdot v=|v| \, \ppsi$, 
and $\pbpsi\cdot \tbpsi=\ppsi\cdot \tpsi=0$.
Note that $\tbpsi,\tpsi$ have $N-1$ independent components, while $\pbpsi,\ppsi$ have one.
With this decomposition, the action \eq{eq:fermionaction} can be rewritten as 
\[
S_{\bpsi\psi}=-\frac{\alpha}{v^2} -N \ln v^2+\tbpsi\cdot \tpsi - \pbpsi\cdot \ppsi-\frac{v^2}{6 \alpha} (\tbpsi \cdot 
\tpsi)^2,
\label{eq:scomp}
\]
where we have used $\pbpsi^2=\ppsi^2=0$. Here we note that the transverse and parallel components are decoupled. 

The parallel component has no interactions, and the integration generates a prefactor $-1$. 

To compute the integral over the transverse components, let us first recall
\s[
\int d\tbpsi d\tpsi\, (\tbpsi\cdot \tpsi)^{2n} e^{\tbpsi\cdot \tpsi}&=\left[\frac{d^{2n}}{d k^{2n}} \int d\tbpsi d\tpsi e^{k \tbpsi\cdot \tpsi} \right]_{k=1}\\
&=\left[\frac{d^{2n}}{d k^{2n}} k^{N-1}  \right]_{k=1}\\
&=(1-N)_{2n}, 
\label{eq:int2n}
\s]
where $(a)_n=a(a+1)\cdots(a+n-1)$ denotes the Pochhammer symbol.
Note that \eq{eq:int2n} are non-zero only if $2n\leq N-1$.
Now let us compute the part with $\tbpsi,\tpsi$ in \eq{eq:trhopsi} with \eq{eq:scomp} by expanding its interaction term:
\s[
\int d\tbpsi d\tpsi  e^{\tbpsi \cdot \tpsi-\frac{v^2}{6 \alpha}(\tbpsi\cdot \tpsi)^{2} } 
&=\sum_{n=0}^\infty \frac{1}{n!} \left(-\frac{v^2}{6 \alpha} \right)^{n} \int d\tbpsi d\tpsi  (\tbpsi\cdot \tpsi)^{2n} e^{\tbpsi \cdot \tpsi} \\
&=\sum_{n=0}^{\lfloor\frac{N-1}{2}\rfloor} \frac{1}{n!} \left(-\frac{v^2}{6 \alpha} \right)^{n} (1-N)_{2n} \\
&=\sum_{n=0}^{\lfloor \frac{N-1}{2}\rfloor}  \frac{1}{n!} \left(-\frac{2 v^2}{3 \alpha} \right)^{n} \left( \frac{1-N}{2} \right)_n \left( \frac{2-N}{2} \right)_n \\
&= \left(\frac{3}{2}\right)^{1 -\frac{N}{2}} \left(\frac{\alpha}{v^2}\right)^{1 -\frac{N}{2}}
 U\left(1 - \frac{N}{2},\frac{3}{2}, \frac{3 \alpha}{2 v^2}\right),
\s]
where $\lfloor \cdot \rfloor$ denotes the floor function, we have used a formula,
\[
(a)_{2n}=2^{2n} \left( \frac{a}{2} \right)_{n} \left( \frac{a+1}{2}\right)_n,
\]
and $U$ is a confluent hypergeometric function of the second kind, which has a relation,
\[
U(a,b,z)=\frac{\Gamma(b-1)}{\Gamma(a)} z^{1-b} {}_1F_1(a-b+1,2-b,z)+\frac{\Gamma(1-b)}{\Gamma(a-b+1)} {}_1F_1(a,b,z),
\]
in terms of the confluent hypergeometric function of the first kind, ${}_1 F_1$.
This finally leads to a compact expression,
\[
\trho(v)=-3^\frac{1}{2} \pi^{-\frac{N}{2}} 2^{-1+\frac{N}{2}}  \alpha\, e^{-\frac{\alpha}{v^2} }  
|v|^{-N-2}
 U\left(1 - \frac{N}{2},\frac{3}{2}, \frac{3 \alpha}{2 v^2}\right).
 \label{eq:general}
\]

\section{Comparisons with numerical simulations}
\begin{figure}
\begin{center}
\includegraphics[width=7cm]{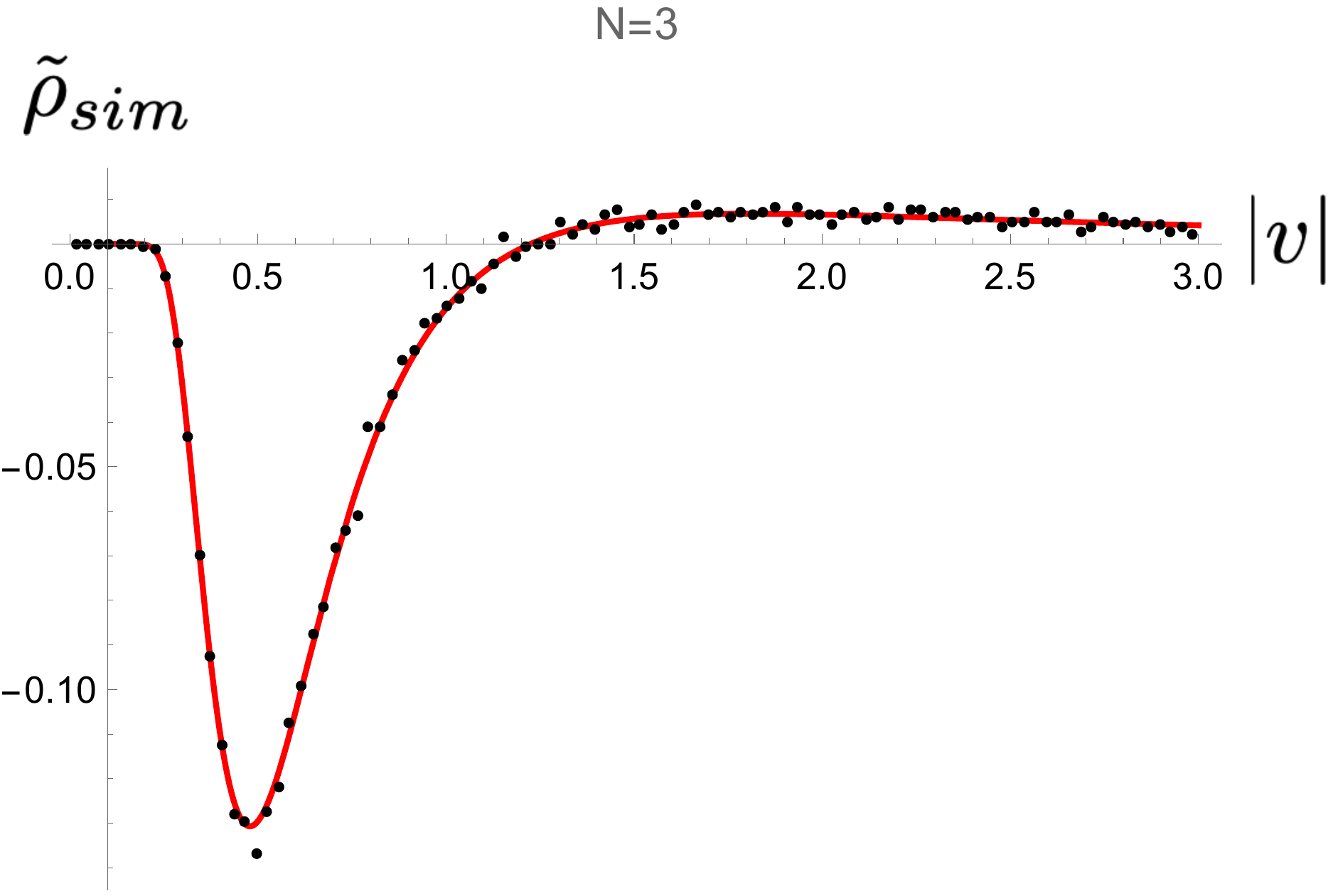}
\hfil
\includegraphics[width=7cm]{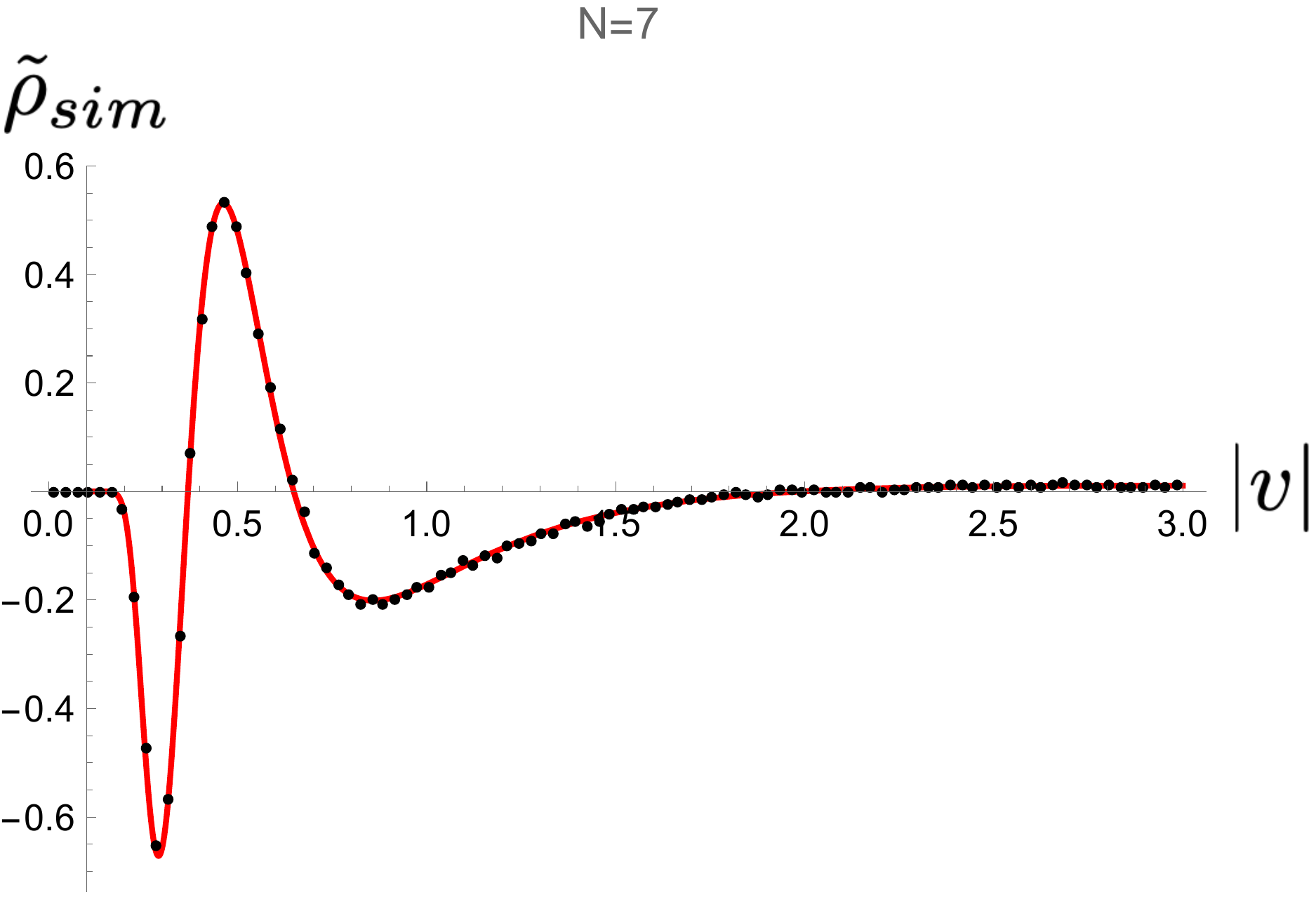}
\caption{The comparison between the numerical values \eq{eq:rhosim} (dots) and the analytical result (solid line)
for $N=3,7$.
The data are generated from $N_C=10000$ random $C$'s with the normal distribution. The bin size is $\delta v=0.03$.}
\label{fig:neq2}
\end{center}
\end{figure}

For a given general value of $C$, one can numerically compute the eigenvectors defined in \eq{eq:cv} by
an appropriate numerical method which solves systems of polynomial equations.\footnote{Such a method generally gives complex solutions as well, and 
one can check whether all the eigenvectors are covered or not by
checking whether the number of these generally complex solutions agrees with the known number, $2^N-1$ \cite{cart}. }
Generally such a method gives complex solutions as well, and we only pick up real eigenvectors to adjust to our interest.
We used Mathematica 12 to solve the equations.
We took the following processes for our numerical simulations.
\begin{itemize}
\item
Generate $C$ by the normal distribution: Each independent component is generated by $C_{ijk}=\sigma/\sqrt{d(i,j,k)}, \ 
(i\leq j \leq k)$. Here $\sigma$ is a random number following the normal distribution of mean value zero and standard 
deviation one. $d(i,j,k)$ is a degeneracy factor defined by 
\s[
d(i,j,k)=\left\{ 
\begin{array}{ll}
1, & i=j=k \\
3 ,& i=j\neq k ,\,i\neq j=k,\, k=i\neq j \\
6 ,& i\neq k \neq j \neq i
\end{array}
\right.
.
\s]
This corresponds to the distribution of $C$  in \eq{eq:defrho} with $\alpha=1/2$,
since  
\s[
C^2=\sum_{i\leq j \leq k=1}^N d(i,j,k) \, C_{ijk} ^2
\s]
due to $C$ being a symmetric tensor.
\item
Compute all the real eigenvectors.
\item
Store the pair of the size $|v|$ and the sign $s$ of $\det M(v)$ for each eigenvector.
\item 
Repeat the above processes.
\end{itemize}
By the above procedure we obtain a data set of $(|v_i|,s_i)\ (i=1,2,\ldots,L)$, where $L$ is the total number of 
data. Then we define
\[
\tilde \rho_{sim}((k+1/2)\delta v)=\frac{1}{N_C}\sum_{i=1}^L s_i \,\theta(k \delta v <|v_i|\leq (k+1)\delta v),
\label{eq:rhosim}
\]
where $N_C$ denotes the total number of randomly generated $C$, $\delta v$ is a bin size, $k=0,1,2,\ldots$, and $\theta$ is a support function which takes 1 if the inequality of the argument
is satisfied, but zero otherwise. 
Then \eq{eq:rhosim} is the numerical quantity corresponding to $\trho(v)S_{N-1} |v|^{N-1} \delta v$ with $\alpha=1/2$, 
because of $\trho(v)dv= \trho(v) S_{N-1} |v|^{N-1} d |v|$, where $S_{N-1}=2 \pi^{N/2}/\Gamma(N/2)$ is
the surface area of the unit hypersphere in $N$-dimensions.
As shown in Figure~\ref{fig:neq2},
we obtain good agreement between the analytical and numerical results. 

\begin{figure}
\begin{center}
\includegraphics[width=7cm]{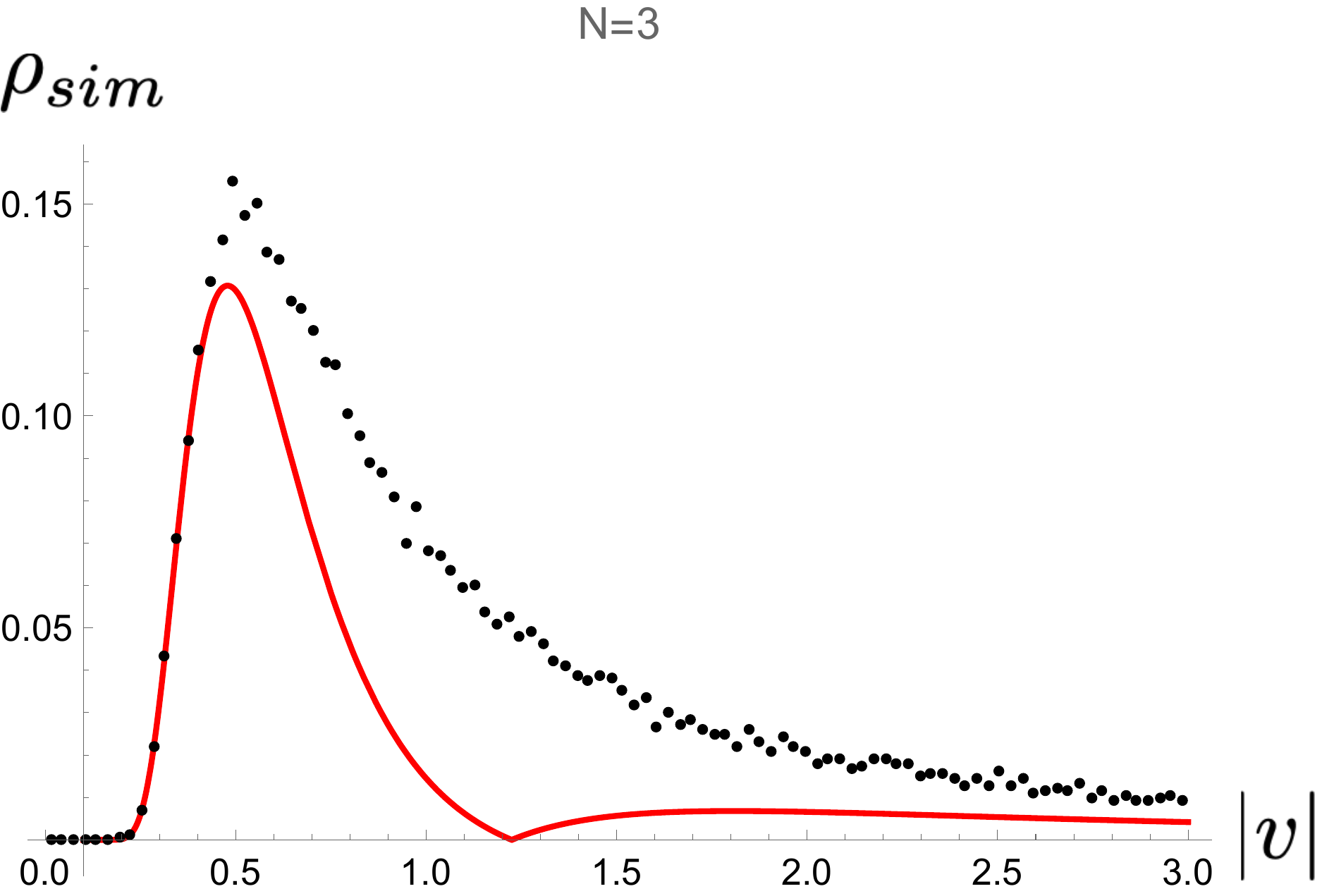}
\hfil
\includegraphics[width=7cm]{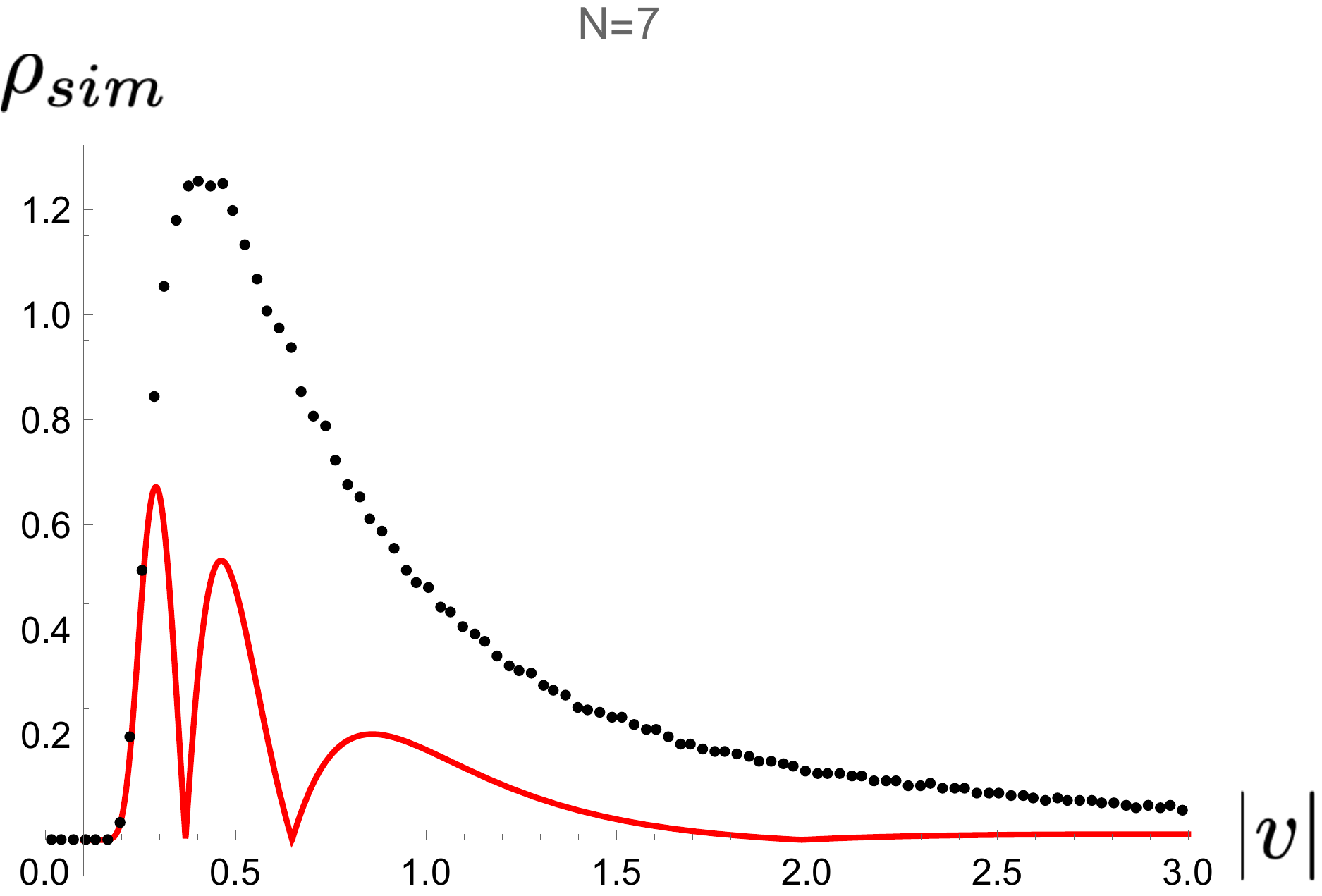}
\caption{
The tightness/looseness of the inequality \eq{eq:ineq} are shown for $N=3,7$.
The dots are \eq{eq:rhosimnos} and the solid line $|\trho(v)| S_{N-1} |v|^{N-1} \delta v$.
The data are generated from $N_C=10000$ random $C$'s with the normal distribution. The bin size is $\delta v=0.03$.}
\label{fig:inequality}
\end{center}
\end{figure}

By ignoring the signs $s_i$ of the same data, we can also consider
\[
\rho_{sim}((k+1/2)\delta v)=\frac{1}{N_C}\sum_{i=1}^L \,\theta(k \delta v <|v_i|\leq (k+1)\delta v),
\label{eq:rhosimnos}
\]
which corresponds to the distribution of real eigenvectors (with no signs). 
As for the inequality \eq{eq:ineq}, the 
bound becomes looser for larger $N$, while it remains tight near the lowest end of the distribution,
as shown in Figure~\ref{fig:inequality}.

\section{Large-$N$ limits}
The formula $\trho(v)$ in \eq{eq:general} has oscillatory behavior and the periods become smaller as $N$ becomes larger. 
Since this is a characteristic structure, we will keep the oscillatory behavior in the large-$N$ limit. 
For this purpose, it turns out that we should perform the following scaling,
\[
\alpha=\frac{\tilde \alpha}{N} 
\]
with fixed $\tilde \alpha$.

With the above scaling, we obtain, for large-$N$,
\[
U\left(1 - \frac{N}{2},\frac{3}{2}, \frac{3 \tilde \alpha}{N 2 v^2}\right)
\sim
\sqrt{2 \pi} e^{-\frac{N}{2}} 2^{-\frac{N}{2}+\frac{1}{4}} N^{\frac{N}{2}} \left(\frac{3 \tilde \alpha}{2 v^2} \right)^{-\frac{1}{4}}
\cdot
\left\{
\begin{array}{ll}
(-1)^{\frac{N}{2}+1} J_{\frac{1}{2}} \left( \frac{\sqrt{3 \tilde \alpha}}{|v|}\right) & \hbox{even }N\\
(-1)^\frac{N-1}{2} J_{-\frac{1}{2}} \left( \frac{\sqrt{3 \tilde \alpha}}{|v|}\right) & \hbox{odd }N
\end{array}
\right.
,
\label{eq:asympu}
\]
where $J$ denotes the Bessel function of the first kind.
To derive this asymptotic expression, we have used the Stirling's approximation and the following properties of 
the hypergeometric functions,
\s[
&\lim_{a\rightarrow \infty} {}_1 F_1 \left(a,b,\frac{z}{a}\right)={}_0 F_1(b,z)=\Gamma(b) (-z)^\frac{1-b}{2} J_{b-1}\left(2 \sqrt{-z}\right).
\s]
From \eq{eq:asympu}, we obtain for $N\rightarrow \infty$, 
\[
\trho(v) \sim -3^{\frac{1}{4}} \pi^{\frac{1-N}{2} }\tilde \alpha^\frac{3}{4} N^{\frac{N}{2}-1} e^{-\frac{N}{2}} |v|^{-N-\frac{3}{2}}  
\cdot
\left\{
\begin{array}{ll}
(-1)^{\frac{N}{2}+1} J_{\frac{1}{2}} \left( \frac{\sqrt{3 \tilde \alpha}}{|v|}\right), & \hbox{even }N\\
(-1)^\frac{N-1}{2} J_{-\frac{1}{2}} \left( \frac{\sqrt{3 \tilde \alpha}}{|v|}\right), & \hbox{odd }N
\end{array}
\right.
.
\]

\section{Summary and future prospects}
In this paper, we have derived an explicit formula for signed distributions of real tensor eigenvectors for 
random real symmetric order-three tensors with Gaussian distributions as the simplest case. 
The formula is expressed in terms of the confluent hypergeometric function of the second kind, which has been derived by 
computing a partition function of a four-fermi theory. We have also discussed the tightness/looseness of the 
formula as lower bounds of real eigenvector distributions, and the large-$N$ limit with the characteristic 
oscillatory behavior of the formula being preserved.

It would be interesting to extend the results in some directions. One is to consider other types of tensors. 
The tensors employed in colored tensor models \cite{Gurau:2009tw} would be an especially interesting case, because of the 
presence of $1/N$ expansions in the models. Another direction would be to consider more complicated distributions 
than Gaussian for tensors, and study the responses to the signed distributions. These extended studies will
reveal the interplays between dynamics and the signed distributions, definining their roles in tensor models. 
It will also be interesting to apply the present result to the spherical $p$-spin model for spin glasses,
using the connection explained in Section~\ref{sec:intro}.

\vspace{.3cm}
\section*{Acknowledgements}
The author would like to thank J.~B.~Geloun for drawing its attention to this interesting problem 
at the conference Group 34, 
and some encouraging comments to a draft.  The author is supported in part by JSPS KAKENHI Grant No.19K03825. 

\appendix
\def\thesection{Appendix \Alph{section}}

\section{Computation of $D^2$}
\label{app:d2}
From \eq{eq:bd}, 
\[
D_a=\left(1+\frac{1}{3\alpha } \bpsi\cdot v \, \psi\cdot v\right) v_a+\frac{1}{3\alpha } \left( \bpsi_a \, \psi\cdot v\, v^2+\bpsi\cdot v \, \psi_a \, v^2 \right).
\]
Therefore,
\s[
D^2&=\left(1+\frac{1}{3\alpha } \bpsi\cdot v \, \psi\cdot v\right)^2 v^2+\frac{4}{3\alpha} \left(1+\frac{1}{3\alpha } \bpsi\cdot v \, \psi\cdot v\right)\bpsi\cdot v \, \psi\cdot v\, v^2 \\
&+\frac{1}{9 \alpha^2} \left( \bpsi_a \, \psi\cdot v\, v^2+\bpsi\cdot v \, \psi_a \, v^2 \right)^2.
\label{eq:d2raw}
\s]
By using $(\bpsi\cdot v)^2=(\psi\cdot v)^2=0$ coming from the fermionic nature of $\bpsi, \psi$, a few terms in 
\eq{eq:d2raw} vanish, and we obtain \eq{eq:d2exp}.

\end{document}